# Discovering new technique for mapping relational database based on semantic web technology

Noreddine GHERABI
Department of Mathematics and Computer Science
Université Hassan 1er, FSTS, LABO LITEN Settat,
gherabi@gmail.com

Khaoula ADDAKIRI
Department of Mathematics and Computer Science,
Université Hassan 1er, FSTS, LABO LITEN Settat,
Morocco

Mohamed BAHAJ
Department of Mathematics and Computer Science
Université Hassan 1er, FSTS, LABO LITEN Settat,
mohamedbahaj@gmail.com

*Abstract*— **Most of data on the Web are still stored in relational databases. Therefore, it is more important to make the correspondence between relational databases (RDB) and ontologies for storing the Web data. In this paper, we present an new approach to map the data stored in relational databases into the Semantic Web, we exploit simple mappings based on some specifications of the database schema, and we explain how relational databases can be used to define a mapping mechanism between relational database and OWL ontology. A framework has been developed, which migrates successfully RDB into OWL document. The experimental results were very important, demonstrating that the proposed method is feasible and efficient.**

## I. INTRODUCTION

The popularity of ontologies is rapidly growing since the emergence of the Semantic Web. in this time, the amount of available Web ontologies continues increasing at a phenomenal rate. However, most of the database today are still locked in data stores and are not published as an open Web of inter-referring resources

The developing Semantic Web ontology languages (e.g., OWL [1]) and services Web (e.g., web mobile agents [2]),it is necessary to use existing data resources, such as relational databases and object-relational databases. Currently, relational databases are the largest data resources in the world, but the structure and integrity constraints of relational tables are defined by schemas, which are not as defined by an ontology Web.

The aim of this paper is to demonstrate a very simple application based on to the transformation of relational database schema to RDF/OWL that is based on correspondence structure between the tables of the database and the classes of the ontology, as well as between table s columns/relationships in the database and datatype/object properties in the ontology.

There is at least a conceptual possibility to create the mapping between the source RDB schema and target OWL ontology by means of model transformations described in some transformation language, bat typically these transformation are not supported on data in RDB or RDF and require the use of an intermediate format.

In this paper, we address the extraction of relational databases by focusing on the conceptual model of the database. It is intended to raise the representation level of the database up to the conceptual model. We developed a prototype (supported by case study) and established an algorithm that enables the mapping process between the database and OWL

The remainder of this paper is organized as follows. Section II discusses some related works. Section III gives an overview of our methodology in details, this section explains our method for extracting the relational schema of database and describes how to transform the relational schema into semantic enrichment model for OWL data (SEMO). After in section IV we implement an algorithm for mapping SEMO to OWL data. Section V evaluates the approach on several cases from real world domains. . Finally, Section VI concludes the paper.

## II. RELATED WORK

Several approaches have been presented that directly map relational schemas to ontology languages [3]. Recently, the W3C RDB2RDF Working Group is developing a direct mapping standard that focuses on translating relational database instances to RDF [4].

At present, many works focus on discovering mapping automatically. For example, Dragut and Lawrence [5] transform respectively the relational schemas and the ontologies into directed labeled graphs. Dou et al. [6] describe a general framework for integrating databases with ontologies via a first-order ontology language Web-PDDL.

With the importance and benefits provided by Web semantic, there has been a lot of effort on migrating RDBs into the relatively newer technologies (XML/RDF/OWL) [7], [8], [9], [10]. Extracting conceptual schema from a logical RDB schema has been also extensively studied [11], [12]. Fonkam et al [13] propose also an algorithm for converting RDB schemas into conceptual models. Cullot et al [14]. use an efficient method for generating classes from tables and converts column to predicate, by using the specific relational database schema characteristics, after the mappings are stored in a R2O document.

## III. OVERVIEW OF THE APPROACH

In this section, we present the proposed process for mapping the RDB into MTR. The process starts by extracting the basic metadata information about a given RDB (See Fig.1).

The first step of our approach consists to extract the structure of the relational database and represent it in a relational Metadata (MTR), after the data of the database is extracted. In the second step we use the Semantic enrichment model for OWL data (SEMO) to facilitate the classification of the tables, column and relationships with cardinalities.

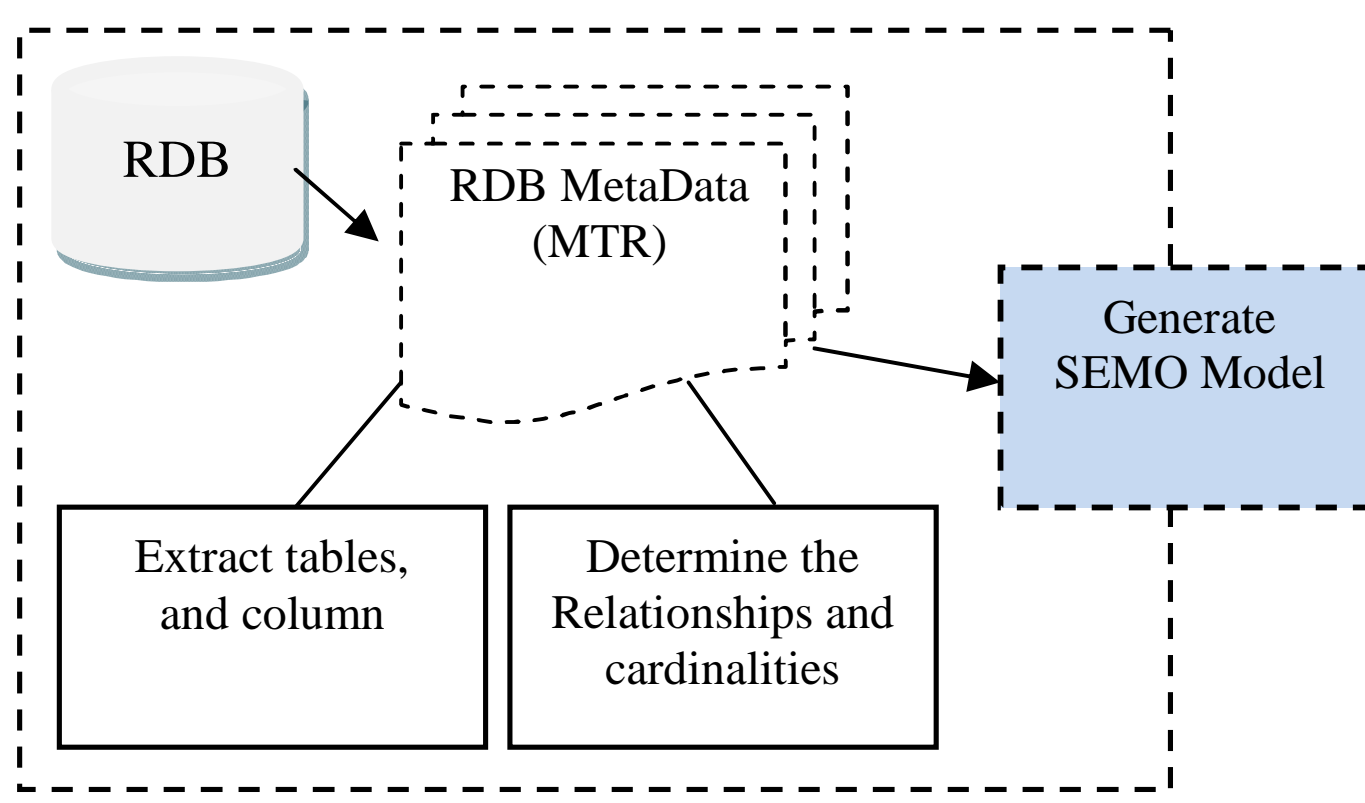

Fig.1: A mapping schema between RDB and MTR

### A. *Our methodology for mapping RDB into MTR*

#### 1) *Extracting MetaData of RDB (MTR).*

In our approach an RDB schema is represented as a set of elements (table name α, set of columns β, Primary Keys γ, Foreign Keys δ and a set of Relationships θ)

$$MTR = \{<\alpha, \beta, \gamma, \delta, \theta>\}$$

- $\alpha$ is the name of the table.
- β describes the set of its column this set is defined as:

$$\beta = \{<\beta_N, \beta_T, \beta_L>\}$$

Where:

- $\beta_N$ is the name of $\beta$.
- $\beta_T$ its type.
- $\beta_L$ is the data length of the column $\beta$.

- $\gamma$ denotes primary key of the table (single valued key or composite key) .
- $\delta$ is the set of foreign key(s) of $\alpha$ , $\delta = \{\delta_N, \delta_R\}$ where $\delta_N$ represents foreign key column name and $\delta_R$ is a name of the second table $\alpha'$ that contains the referenced $\delta$ .
- Relationships $\theta$ : A table $\alpha$ has a set of relationships $\theta$ .

The kinds of RDB tables and relationships are identified using primary/foreign keys, Each relationship (R϶ $\theta$ ) between a table $\alpha$ and another table $\alpha'$ is defined as:

$$\theta(\alpha, \alpha') := \{R \mid R := (\gamma(\alpha), \alpha, \delta(\alpha), \alpha', C)\}$$

Where $\gamma(\alpha)$ is the primary key of $\alpha$ , $\delta(\alpha)$ is the foreign key representing the relationship in $\alpha'$ and C the cardinality of the source table $\alpha$ .

#### 2) *Algorithm for extraction of MTR*

This section presents the algorithm for extracting *MTR*, is used to extract the information about MetaData of RDB, which contains the names of the tables, columns and integrity constraints of all the tables extracted from an RDB.

The algorithm for extraction the MTR from RDB is as follows:

Algorithm MTR (BD: RDB) return MTR

MTR: = null; // a set to store RDB

For each table T ϵ RDB do

$\alpha_N$ := Extract name of ( T)

$\beta_N$ :=ExtractClumn(T)

$\beta_T$ :=ExtractColumnType(T)

$\beta_L$ :=Extractlengthofthecolumn (T)

End For

$\gamma$ :=ExtractPrimaryKeys (T)

$\delta$ := ExtractForeignKeys(T)

End For

For each set of tables (T, T') Create element $\gamma$ for storing the prosperities of the relationships between T and T'.

$\gamma(\alpha)$:= ExtractPrimaryKey (T)

$\alpha$ := ExtractTable(T)

$\delta(\alpha)$:= ExtractForeignKey(T')

```
α' := ExtractTable (T')
C:= ExtractCardinalitySource(T)
End For
MTR: = MTR+ T // add the table T to MTR
Return MTR
 End algorithm
```

*B. Semantic enrichment model in OWL(SEMO)*

Semantic enrichment model (SEMO) in semantic Web is a process of analyzing and examining a database to capture its structure and definitions at a higher level of meaning.
The SEMO retains all data semantics that could be extracted from an RDB and the integrity constraints imposed on it. However, it acts as a key mediator for converting existing RDB data into target databases based on the structure and the concepts of the target models
The next step is to define the SEMO based on a classification of tables, columns and relationships, which may be performed through data access.
To convert a MTR into SEMO Classes, we define a class OWL as root and put its relevant child into a document. We load MTR model into object instances in the Web ontology.
The SEMO Model is defined as a set of classes, is denoted as 3-tuple:

$$SEMO := \{Owl - C_N, Owl - C_A, Owl - C_R\}$$

Where the first element is the name of the class SEMO, the second element is a list of attributes and the latest element is the relationships between classes

The attributes defined as:

$$Owl - C_A := \{Owl - A_N, Owl - A_T\}$$

Where $Owl - A_N$ is an attribute name, $Owl - A_T$ is its type.

$Owl - C_R$ describes the different types of relations that can exist between any pair of classes in the CDM.

$$Owl - C_R := \{Owl - R_s,, Owl - R_d, Owl - R_c\}$$

Where $Owl - R_s$ is the source class and $Owl - R_d$ is the destination class and $Owl - R_c$ is the Cardinality source of the class.

## IV. OWL STRUCTURE

In this section we describe the mapping process for generating the structure of OWL document from SEMO model. the process for generating is illustrated in figure 2.
When the SEMO Model has been obtained, the specified rules are used to map the SEMO constructs into OWL classes and create nodes for storing OWL data

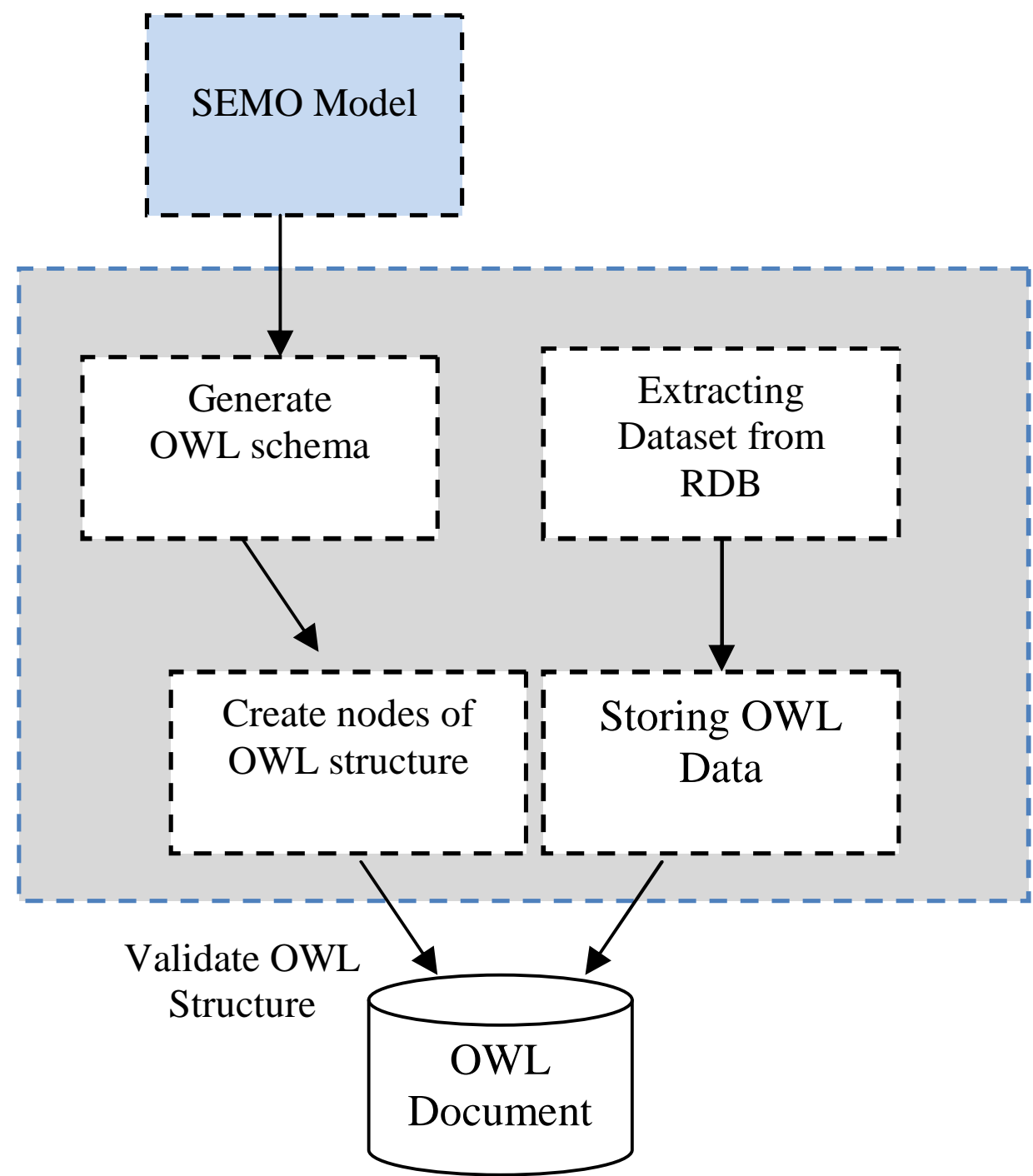

Fig.2: Architecture for generating Owl data from SEMO model.

Each class in OWL is structured by a group of individuals. Every individual in the OWL is a subclass of the main class owl:Thing. (Fig.3)

The class in SEMO Model is translated to owl:class in the Web ontology, our class in OWL technology is represented as follows:

$$< \text{owl : Class rdf : ID ="} \text{C} \in Owl - C_N \text{" / >}$$

Each attribute A is translated into a owl:DatatypeProperty class and represented as :

$$< \text{owl : DataTypePr operty rdf : ID ="A} \in Owl - A_N \text{">}$$
$$< \text{rdfs : domain rdf : resource ="\#C} \in Owl - C_N \text{" / >}$$
$$< \text{rdfs : range rdf : resource ="\&xsd, Type} \in Owl - A_T \text{" / >}$$
$$< \text{/owl : DatatypePr operty >}$$

The relationship between two classes C1 and C2, the representation of the relationship in Web ontology is represented as follows:

$$< \text{owl : ObjectProp erty rdf : ID ="Relation} \in OWL - C_R \text{">}$$
$$< \text{rdfs : domain rdf : resource ="\#} C_1 \in Owl - R_s \text{" / >}$$
$$< \text{rdfs : range rdf : resource ="\#} C_2 \in Owl - R_d \text{" / >}$$
$$< \text{/owl : ObjectProp erty >}$$

The cardinalities of a relationship are given by specifying minimum and maximum cardinalities.

For mapping the general cardinality we use:

<owl:Cardinality rdf:datatype=”&xsd,nonNegativeInteger”> Cardinality ϵOwl-$R_c$</ owl:Cardinality>

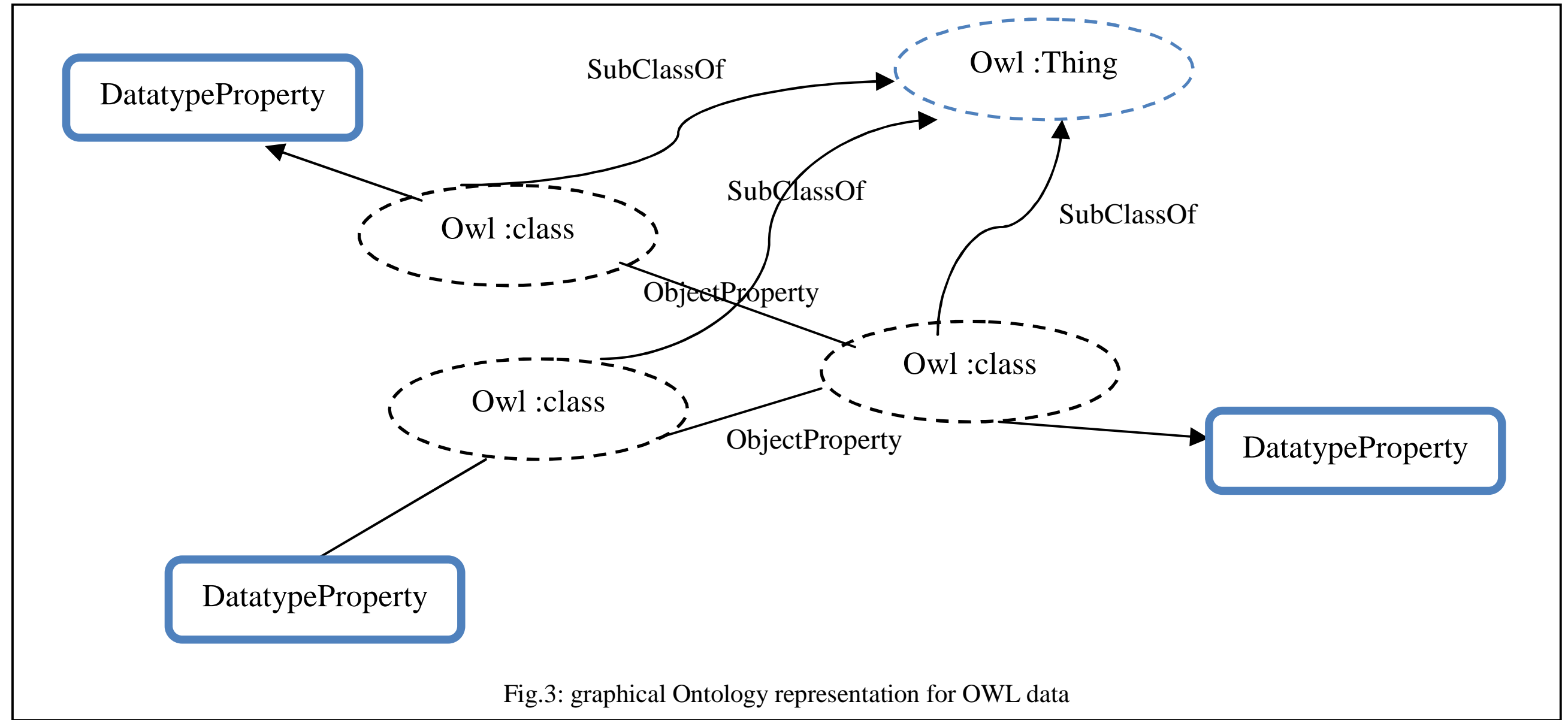

Fig.3: graphical Ontology representation for OWL data

## V. EVALUATION

We have implemented the proposed approach in Java. In this section, we report on some results of an experimental study. The database management system (DBMS) used in this paper is Oracle 11g.

Figure 4 shows a simple example of relational database:

Our system converts the existing RDB data to the SEMO model using the classification of the Relational schema Figure 5 illustrates the SEMO model exported by the system.

After the SEMO model is mapping into complex individuals in OWL document. All tables, column and relationships are mapped into classes in Owl structure. The platform in figure 6 shows the final structure of OWL document, this structure defines the ontology of our example.

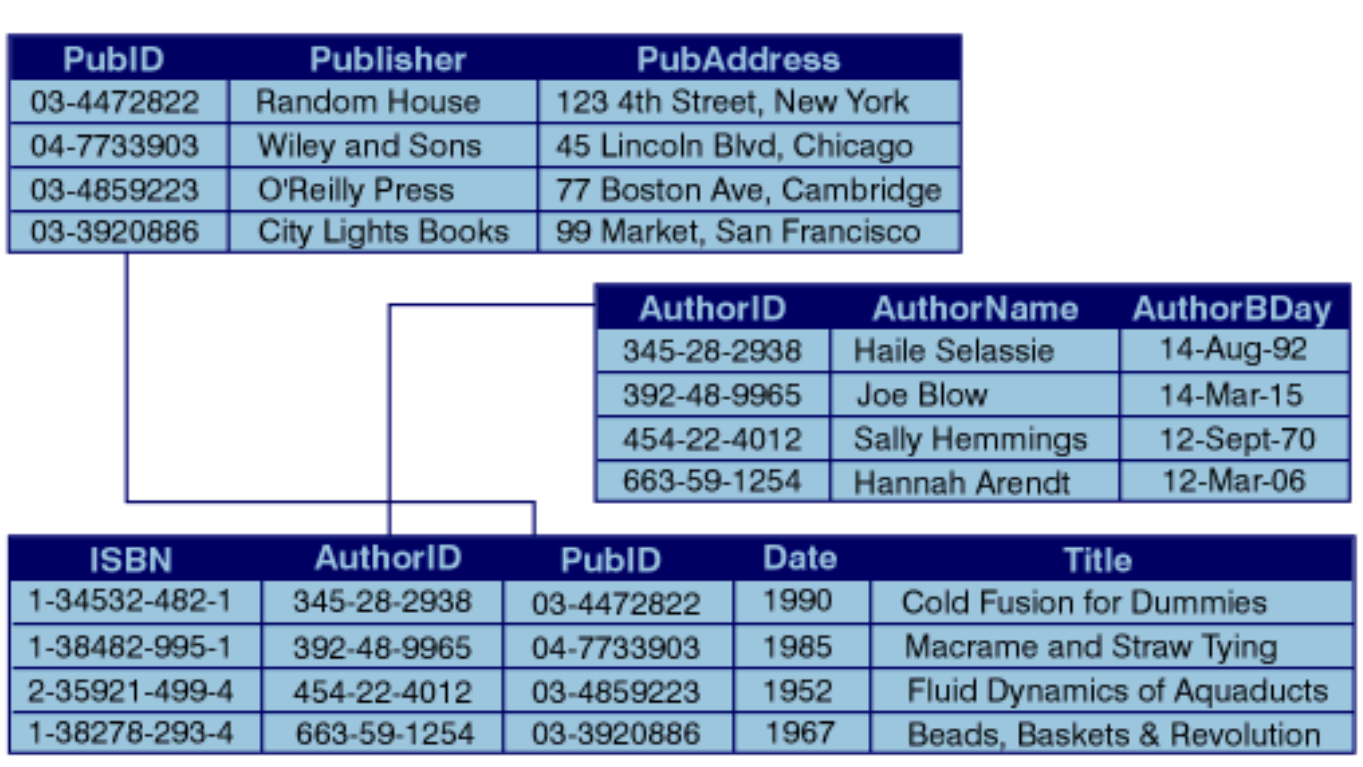

| PubID | Publisher | PubAddress |
|---|---|---|
| 03-4472822 | Random House | 123 4th Street, New York |
| 04-7733903 | Wiley and Sons | 45 Lincoln Blvd, Chicago |
| 03-4859223 | O'Reilly Press | 77 Boston Ave, Cambridge |
| 03-3920886 | City Lights Books | 99 Market, San Francisco |

| AuthorID | AuthorName | AuthorBDay |
|---|---|---|
| 345-28-2938 | Haile Selassie | 14-Aug-92 |
| 392-48-9965 | Joe Blow | 14-Mar-15 |
| 454-22-4012 | Sally Hemmings | 12-Sept-70 |
| 663-59-1254 | Hannah Arendt | 12-Mar-06 |

| ISBN | AuthorID | PubID | Date | Title |
|---|---|---|---|---|
| 1-34532-482-1 | 345-28-2938 | 03-4472822 | 1990 | Cold Fusion for Dummies |
| 1-38482-995-1 | 392-48-9965 | 04-7733903 | 1985 | Macrame and Straw Tying |
| 2-35921-499-4 | 454-22-4012 | 03-4859223 | 1952 | Fluid Dynamics of Aquaducts |
| 1-38278-293-4 | 663-59-1254 | 03-3920886 | 1967 | Beads, Baskets & Revolution |

Fig.4: Example of RDB

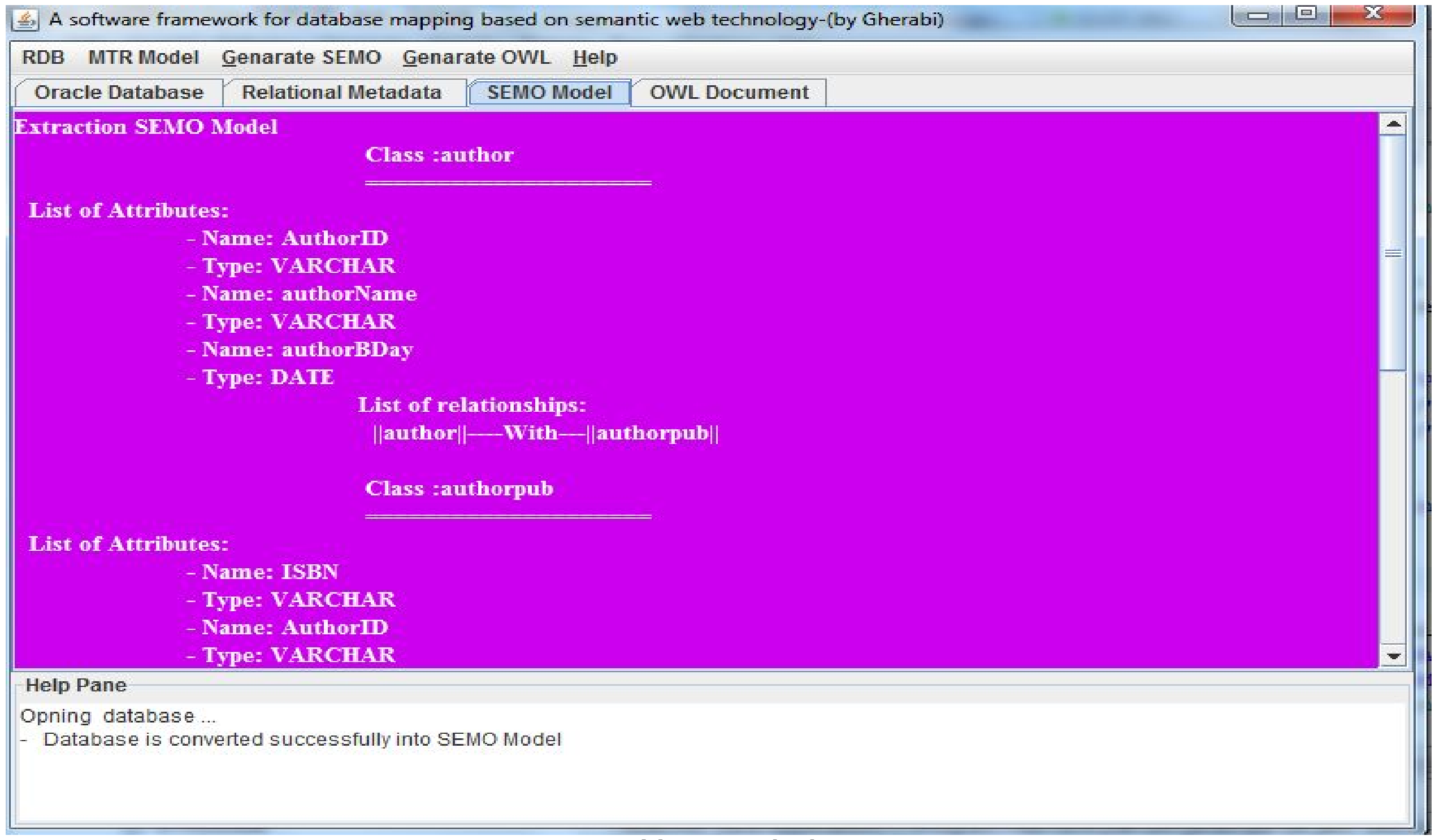

Fig.5: SEMO model extracting by the our prototype

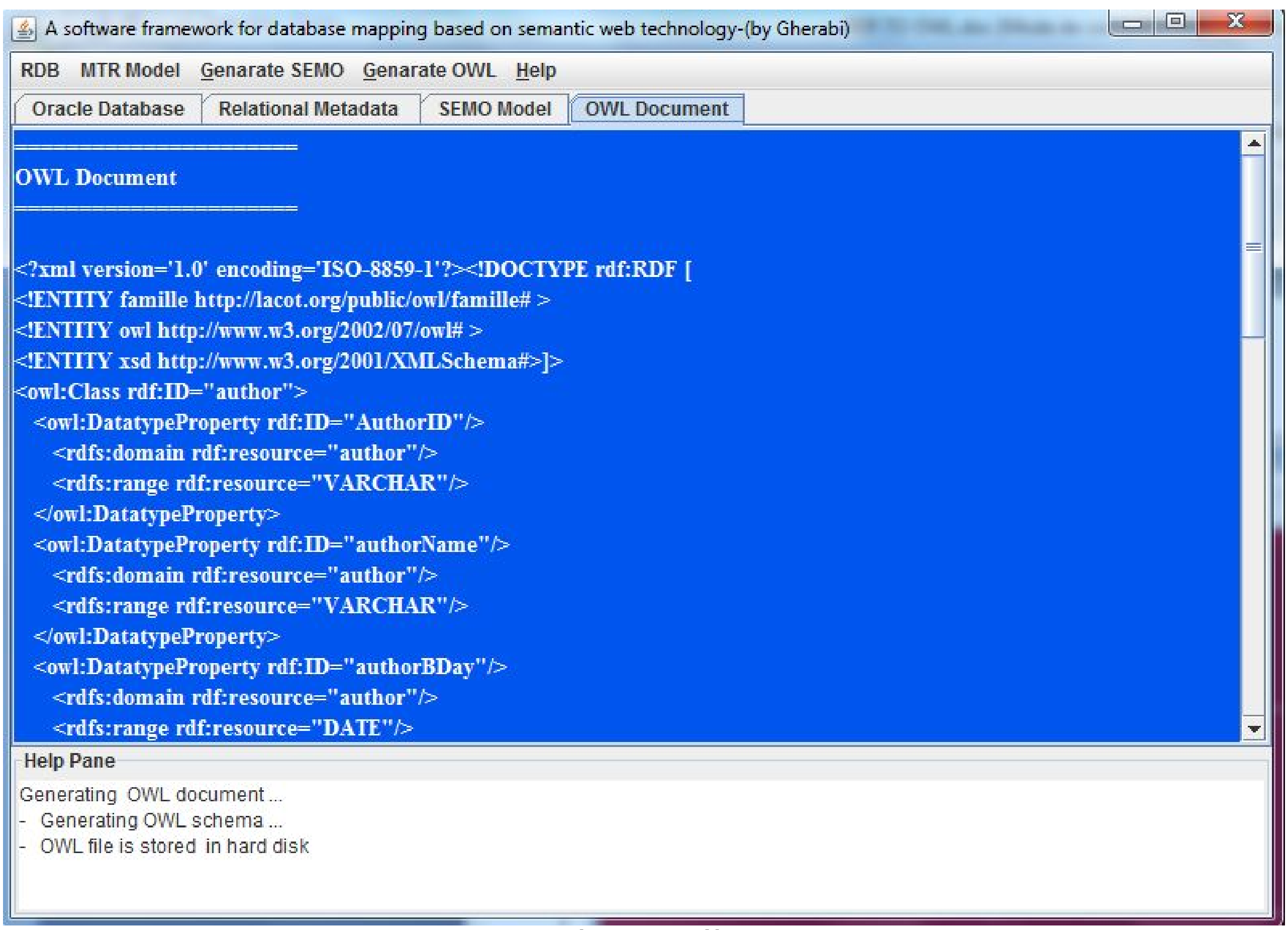

Fig.6: OWL schema extracted by our system

Once our OWL document is written, it's a good idea to ensure the validity and the consistency of its concepts. In a more general observance of a standard or definition of a format promotes interoperability by allowing the developer to ensure the integrity of the data contained in the document OWL. In this paper we use the W3C validator [15], it is used to validate RDF documents. It can therefore also ensure that an OWL document follows the syntax of RDF

## VI. CONCLUSION

In this paper we have presented a fully automated approach to converting the relational database into OWL ontologies, with the respect of the logical rules of passage. We first present the mathematical formulation of our approach, after de the schema of relational database is extracted with constraints and our model called Semantic Enrichment Model of OWL (SEMO) captures the classification of the data structure and an efficient algorithm generates the schema of OWL data.

We have experimentally approved our method on some database from real world domains. The results demonstrate that our approach performs well as compared to some existing methods.

## VII. REFERENCES

[1] OWL Web Ontology Language. http://www.w3.org/TR/owl-ref/

[2] S. A. McIlraith and D. L. Martin. Bringing Semantics to Web Services. IEEE Intelligent Systems, 18(1):90–93, 2003.

[3] W. J. Pardi, XML in Action, Microsoft Press, Washington, 1999.

[4] Fahrner, C. and Vossen, G.: Transforming Relational Database Schemas into Object-Oriented Schemas According to ODMG-93.In 4th Int. Conf. on Deductive and Object-Oriented Databases,pp. 429–446, 1995.

[5] Dragut, E., Lawrence, R.: Composing mappings between schemas using a reference ontology. In Proceedings of International Conference on Ontologies, Databases and Applications of Semantics (ODBASE'04). (2004) 783-800

[6] Dou, D., LePendu, P., Kim, S., Qi, P.: Integrating databases into the Semantic Web through an ontology-based framework. In Proceedings of the 3rd International Workshop on Semantic Web and Databases (SWDB'06). (2006)

[7] Green, J., Dolbear, C., Hart, G., Engelbrecht, P., Goodwin, J."Creating a semantic integration system using spatial data", , in *International Semantic Web Conference 2008* Karlsruhe, Germany

[8] Noreddine Gherabi and Mohamed Bahaj. Robust Representation for Conversion UML Class into XML Document using DOM.

*International Journal of Computer Applications* 33(9):22-29, November 2011

[9] Cristian P´erez de Laborda and Stefan Conrad. Relational.OWL - A Data and Schema Representation Format Based on OWL. In Second Asia-Pacific Conference on Conceptual Modelling (APCCM2005), volume 43 of CRPIT, pages 89–96, Newcastle, Australia, 2005.

[10] Tirmizi et al, "Translating SQL Applications to the Semantic Web", Tirmizi, S., Sequeda, J., Miranker, D., Lecture Notes in Computer Science, Volume 5181/2008 Database and Expert Systems Applications-(2008)

[11] Wu, Z., Chen, H., Wang, H., Wang, Y., Mao, Y., Tang, J., Zhou, C., "Dartgrid: a Semantic Web Toolkit for Integrating Heterogeneous Relational Databases", Semantic Web Challenge at 4th International Semantic Web Conference (ISWC 2006), Athens, USA, 5-9 November 2006.

[12] Alhajj, R.: Extracting the Extended Entity-Relationship Model from a Legacy Relational Database. Inf. Syst, vol. 28, pp. 597–618, 2003.

[13] Fonkam, M. M. and Gray, W. A.: An Approach to Eliciting the Semantics of Relational Databases. In 4th Int. Conf. on Advanced Info. Syst. Eng., vol. 593, pp. 463–480, 1992.

[14] Cullot, N., Ghawi, R., Yetongnon, K..,"DB2OWL: A Tool for Automatic Database to Ontology Mapping", In Proc. of 15th Italian Symposium on Advanced Database Systems (SEBD 2007), pages 491-494, Torre Canne, Italy, 17-20 June 2007.

[15] http://www.w3.org/RDF/Validator/